\input{aipcheck}

\documentclass[
    ,final            % use final for the camera ready runs
  ]
  {aipproc}

\layoutstyle{6x9}
\renewcommand\XFMtitleblock{%
  \XFMtitle
  \let\XFMoldpar\par
  \def\par{\XFMoldpar\def\par{\space
             (on behalf of the H.E.S.S. Collaboration)\XFMoldpar}}%
   \XFMauthors
   \let\par\XFMoldpar
   \XFMaddresses
   \XFMabstract
   \vspace{5pt}%
   \XFMkeywords
   \XFMclassification
}
\begin{document}

\title{Determining atmospheric aerosol content with an infra-red radiometer}
% general physics
%07.20.Ka 	High-temperature instrumentation; pyrometers 
%07.57.Kp 	Bolometers; infrared, submillimeter wave, microwave, and 
%		radiowave receivers and detectors 
%               (see also 85.60.Gz Photodetectors in electronic and magnetic 
%		devices, and 95.55.Rg Photoconductors and bolometers in 
%		astronomy)
%07.60.Dq 	Photometers, radiometers, and colorimeters 
% astronomy & geophysics
%92.60.-e 	Properties and dynamics of the atmosphere; meteorology 
%		(see also 92.40.Zg Hydrometeorology, hydroclimatology)
%92.60.Fm 	Boundary layer structure and processes
%92.60.H- 	Atmospheric composition, structure, and properties 
%92.60.hf 	Tropospheric composition and chemistry, constituent transport and chemistry
%92.60.hv 	Pressure, density, and temperature
%92.60.Jq 	Water in the atmosphere 
%92.60.Kc 	Land/atmosphere interactions 
%92.60.Mt 	Particles and aerosols
%92.60.Wc 	Weather analysis and prediction 
%95.55.-n 	Astronomical and space-research instrumentation
%95.55.Aq 	Charge-coupled devices, image detectors, and IR detector arrays
%95.55.Ka 	X- and \u03b3-ray telescopes and instrumentation 
%95.55.Vj 	Neutrino, muon, pion, and other elementary particle detectors; cosmic ray detectors 
%95.85.-e 	Astronomical observations (additional primary heading(s) must be chosen with these entries to represent the astronomical objects and/or properties studied)
%95.85.Pw 	\u03b3-ray 
\classification{07.57.Kp, 92.60.Fm, 92.60.Mt, 95.55.Ka, 95.55.Rg, 95.85Pw}
\keywords      {lidar, infrared instrumentation, ground based gamma-ray astronomy}

\author{M.~K.~Daniel}{
  address={Department of Physics, University of Durham, Durham, DH1 3LE. U.K.}
}

\author{G.~Vasileiadis}{
  address={LUPM, Un. Montpellier II CC-072, Place Eugenie Bataillon, 34095 Montpellier, France.}
}

%\author{the H.E.S.S. Collaboration}{
%  address={http://www.mpi-hd.mpg.de/hfm/HESS/HESS.shtml}
%  %,altaddress={<author1 address>} % additional visiting address
%}

\begin{abstract}
   The attenuation of atmospheric Cherenkov photons is
   dominated by two processes: Rayleigh scattering from
   the molecular component and Mie scattering from the
   aerosol component. Aerosols are expected to contribute
   up to 30 Wm$^{-2}$ to the emission profile of the atmosphere,
   equivalent to a difference of $\sim20^\circ$C to the clear sky
   brightness temperature under normal conditions. Here
   we investigate the aerosol contribution of the measured
   sky brightness temperature at the H.E.S.S. site; compare
   it to effective changes in the telescope trigger rates; and
   discuss how it can be used to provide an assessment of
   sky clarity that is unambiguously free of telescope systematics.
\end{abstract}

\maketitle

%%%%%%%%%%%%%%%%%%%%%%%%%%%%%%%%%%%%%%%%%%%%
%% MAINMATTER
%%%%%%%%%%%%%%%%%%%%%%%%%%%%%%%%%%%%%%%%%%%%

\section{Introduction}
The atmosphere is the most important part of the detector in ground-based
gamma-ray astronomy, but it is also the part that has the greatest systematic
uncertainty and over which we have the least control. It falls upon us to 
instead monitor and characterise the atmospheric conditions at the time of 
observations so that we can either feed this information into Monte Carlo 
simulations or reject data when conditions go out of acceptable parameters.

After being generated in the upper atmosphere Cherenkov light will either reach 
the ground or be attenuated through the process of Rayleigh scattering on the 
molecular component of the atmosphere, or Mie scattering on the aerosol 
component (variously dust. silicates, pollens, etc). The molecular component 
tends to change relativiely slowly, through seasonal variations; whereas the 
aerosol component can change more rapidly, depending on eg wind conditions. It 
becomes vitally important to characterise this aerosol component of the 
atmosphere through regular monitoring. A lidar is generally used to measure the 
atmospheric transmission (eg \cite{Nolan}) from backscattered laser light. At 
the H.E.S.S. site a lidar centred at 355 and 532nm has been running in 
conjunction with observations since mid-2011. Whilst lidars are excellent 
instruments for determining the presence of aerosols they are not without
complications. Firstly a lidar, due to geometric viewing considerations, only 
becomes effective above a minimum altitude. Secondly, in order to obtain a 
transmission profile relevant to the Cherenkov spectrum the laser wavelengths 
are close to the peak in the emission, this means the lidar is operated only
inbetween observing runs to avoid any light contamination to the telescope 
images. In this paper we look at utilising another piece of the H.E.S.S. 
atmospheric monitoring equipment to fill in some of this missing information.

The atmosphere is split into regions according to its temperature behaviour. 
The troposphere is the lowest, most dense, part of the atmosphere -- where 
most of the weather happens -- and is characterised by a linear decline in 
temperature with increasing altitude and vertical mixing. The molecular density 
profile falls off exponentially, with a scale height of a ~few km; the vertical 
air motion in this region mixes in the larger aerosols which have a smaller 
scale height of order a ~km. The molecular component is an inefficient 
black-body radiator in the 8-14$\mu$m region of the spectrum, water vapour and 
aerosols are slightly more efficient and clouds are very efficient. This makes 
an infra-red radiometer an effective cloud monitor, with clouds showing up as 
a large brightness temperature compared to a relatively ``cold" sky 
\cite{Buckley}. H.E.S.S. employ Heitronics KT19.82 radiometers with 2$^\circ$
field of view to monitor for the presence of clouds, with each telescope having 
a paraxially mounted unit and a further one continuosly scanning the whole sky. 
The infra-red luminosity of the sky ($L_{\mathrm{sky}}$) is a collective sum of 
the emission of a number of different constituent parts
\begin{equation}
L_{\mathrm{sky}} = \epsilon_{l}\sigma T_{\mathrm{lens}}^4
                 + \epsilon_{wv}\sigma T_{wv}^4
                 + \epsilon_{a}\sigma T_{a}^4
                 + \epsilon_{m}\sigma T_{m}^4
                 + \ldots
\end{equation}
where $\epsilon$ is the emissivity of the lens ($l$) and the water
vapour $wv$, the aerosols $a$, and the molecular ($m$) profiles of the
atmosphere, etc and T is the relevant integrated temperature profile in the 
line of sight. According to \cite{Dalrymple} the aerosol component can 
contribute up to 30Wm$^{-2}$ to the bolometric luminosity, which can mean the 
difference between a brightness temperature of -56$^\circ$C or -70$^\circ$C in 
the presence or absence of aerosols respectively. This leads to the prospect of 
changing aerosol conditions leading to a noticeable change in the sky brightness
temperature ($T_{\mathrm{sky}}$) measurements.

\section{Data and Observations}
\begin{figure}[p]
  \includegraphics[height=.25\textheight]{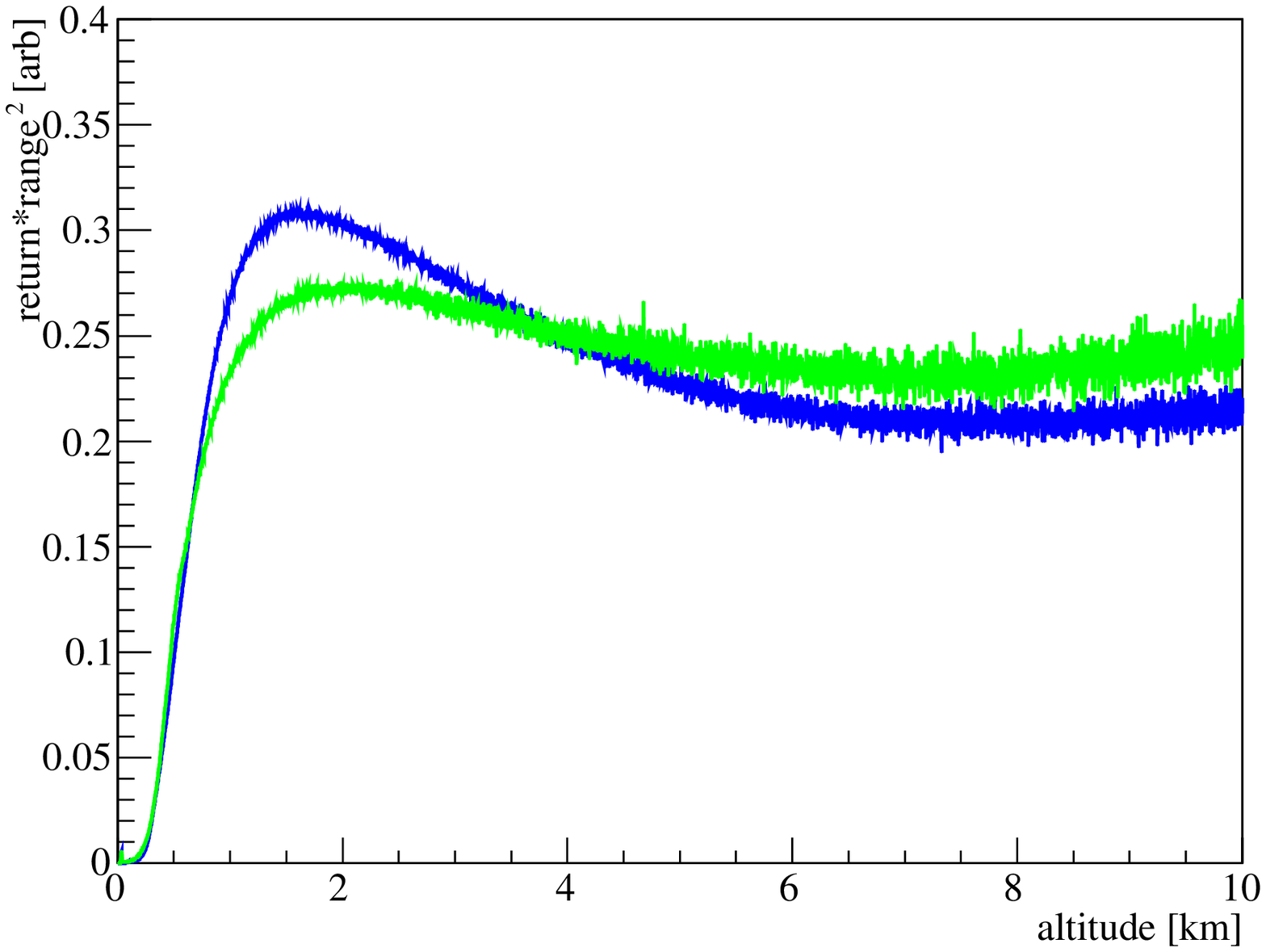}
  \includegraphics[height=.25\textheight]{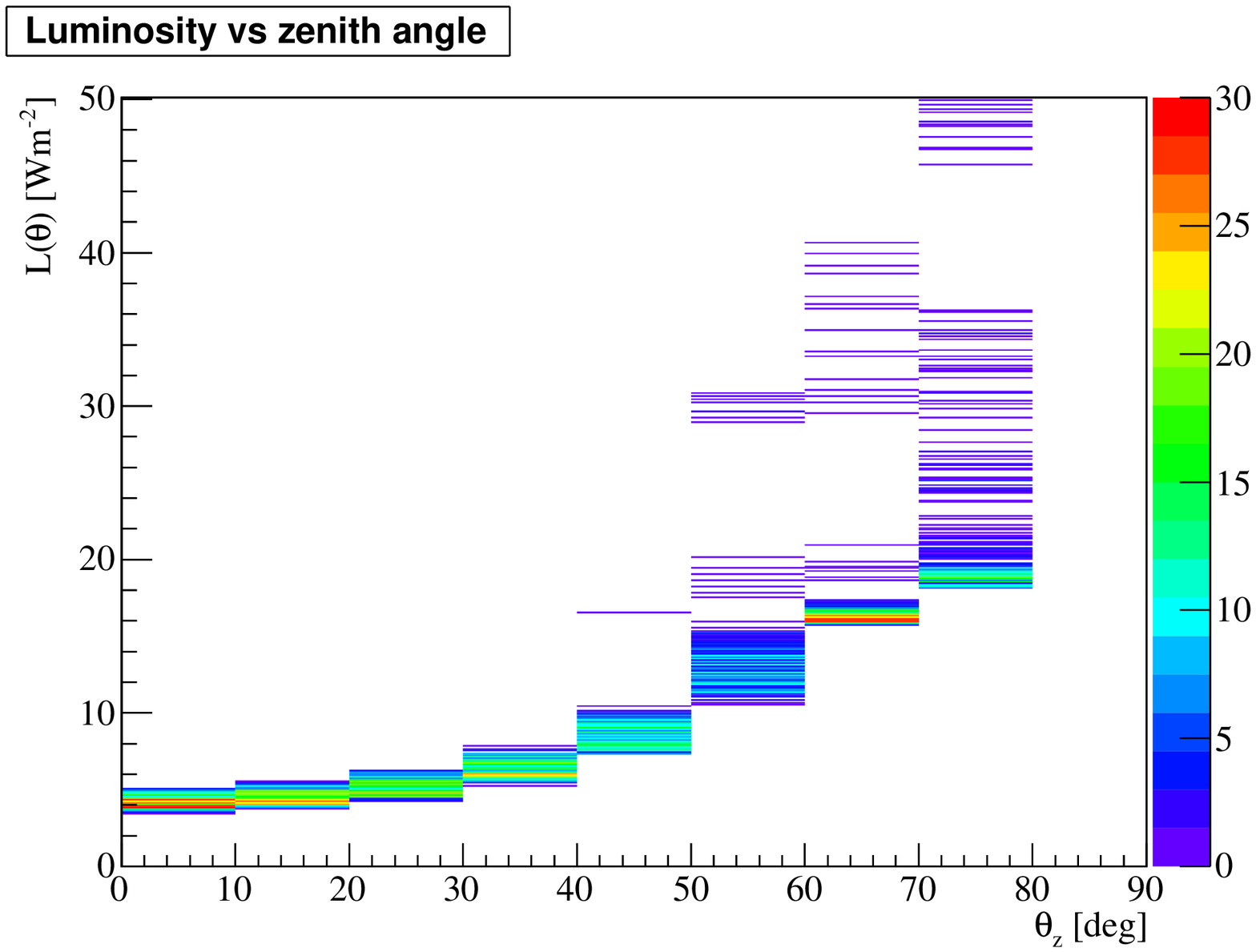}
  \caption{Sky conditions on the night of 20/08/2011. The left hand plot is the
  lidar profile (blue at 355nm, green at 532nm), showing no observable aerosol 
  boundary layer; the right plot gives the histogram of infra-red luminosity 
  measurements as a function of zenith angle.}
  \label{fig:20110820}
\end{figure}

\begin{figure}[p]
  \includegraphics[height=.25\textheight]{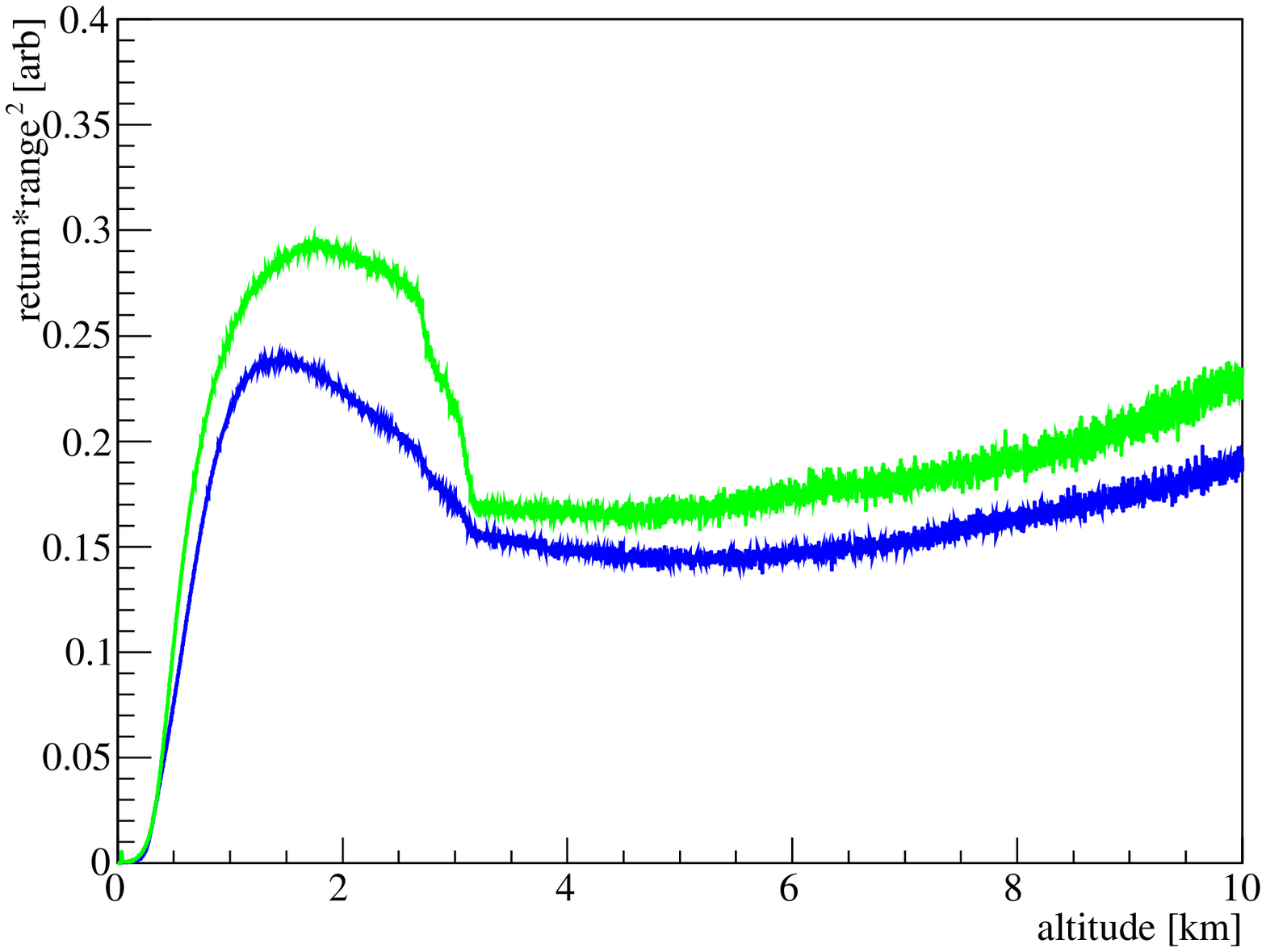}
  \includegraphics[height=.25\textheight]{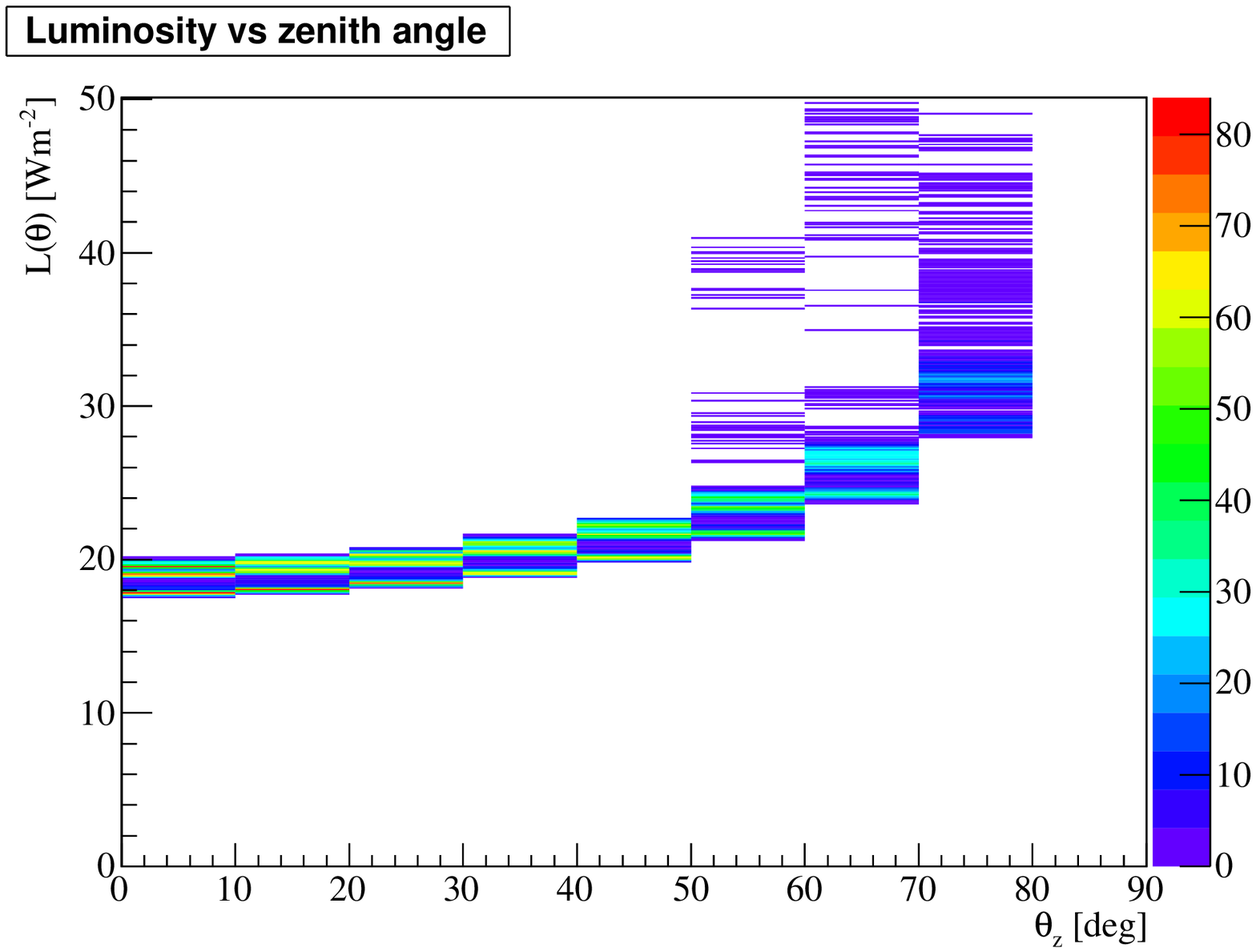}
  \caption{As figure~\ref{fig:20110820} but for the night of 29/08/2011. There
  is a prominent aerosol component up to a boundary layer of $\sim3$km and the
  infra-red lumonisity is substantially increased.}
  \label{fig:20110829}
\end{figure}

\begin{figure}[p]
  \includegraphics[height=.25\textheight]{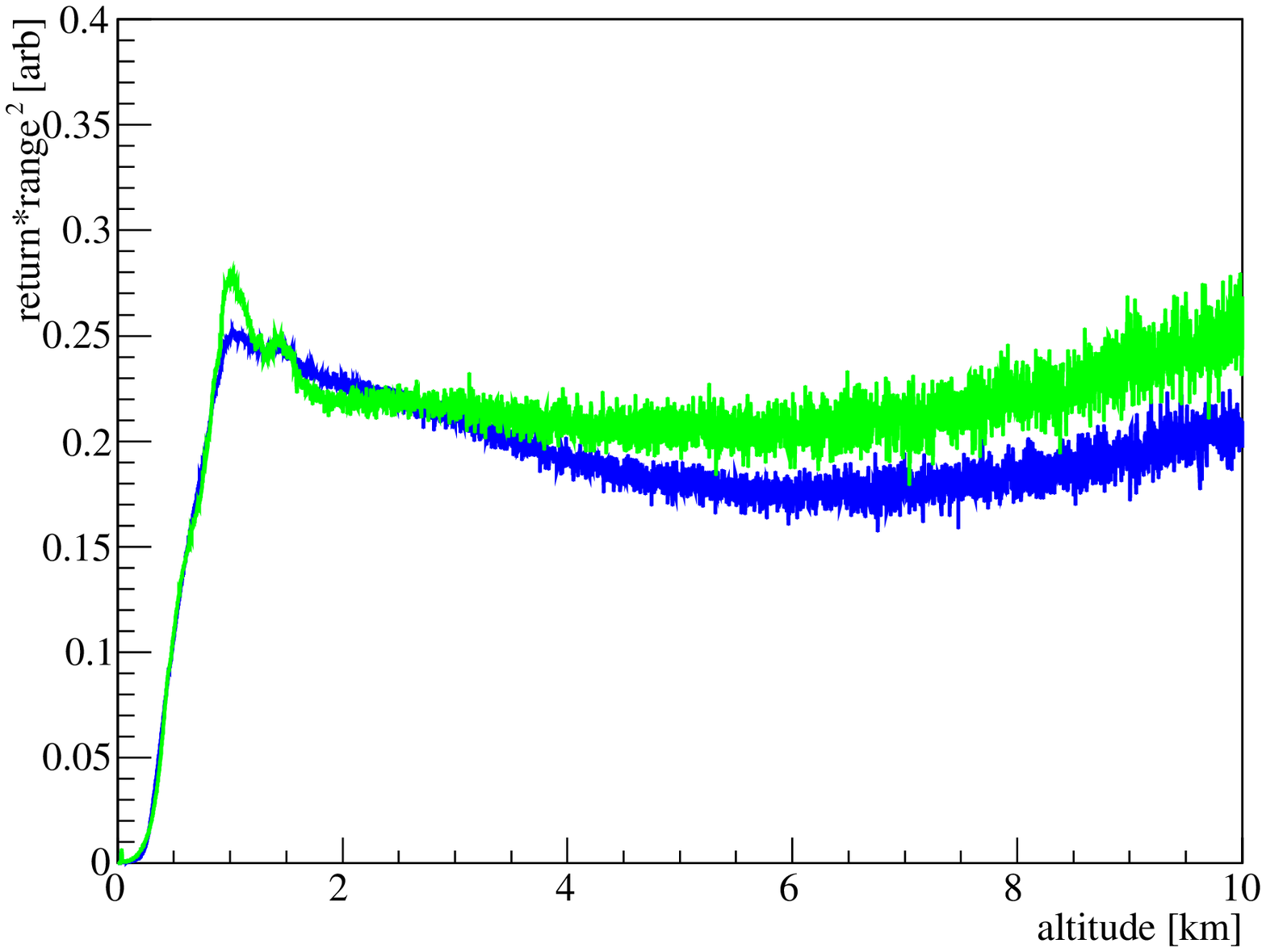}
  \includegraphics[height=.25\textheight]{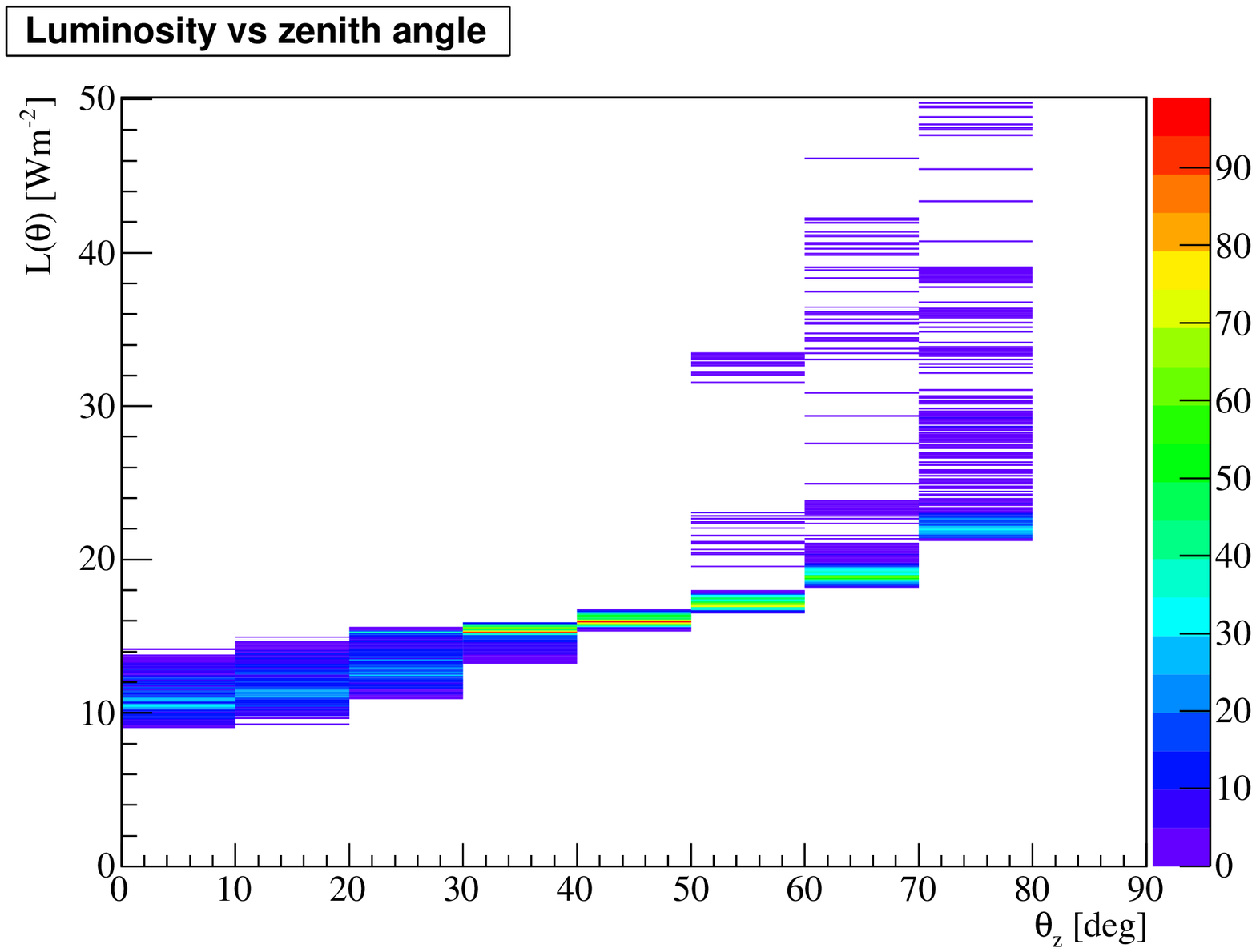}
  \caption{As figure~\ref{fig:20110820} but for the night of 01/09/2011. There
  is a noticeable aerosol component up to a boundary layer of $\sim1$km and the
  infra-red lumonisity is moderately increased.}
  \label{fig:20110901}
\end{figure}

The August to September period at the H.E.S.S. site often has noticeable aerosol 
contamination due to biomass burning in neighbouring countries and the resultant 
smoke being blown downwind. In figure~\ref{fig:20110820} we see an ``ideal'' 
night which has no measurable aerosol contribution (the large particles having 
sedimented out of the atmosphere); within the space of a week 
figure~\ref{fig:20110829} shows ``hazy'' conditions, with a prominent aerosol 
boundary layer that extends up to about $\sim 3$km; a couple of days later 
figure~\ref{fig:20110901} shows the aerosols sedimenting out once more, with the
boundary layer close to the lidar effective altitude threshold at $\sim 1$\,km 
(characteristic of ``normal'' observing conditions).

\begin{figure}[htb]
  \includegraphics[height=0.25\textheight]{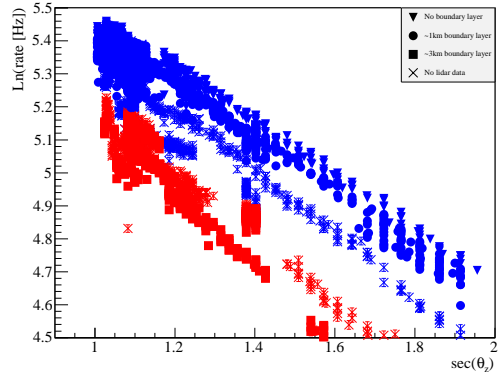}
  \caption{Telescope trigger rates as a function of zenith angle. 
   The triangles correspond to periods with no observed boundary layer,
   circles when the boundary layer is at $\leq 1$km, 
   squares when the boundary layer reaches $\sim 3$km and crosses for when 
   there are no measurements available. The red points are when $T_\mathrm{sky}$
   at zenith is $\geq -50^\circ$C, blue points when it is lower.}
  \label{fig:rates}
\end{figure}

In figure~\ref{fig:rates} we show the telescope trigger rates as a function of
zenith angle for all observing runs for that osberving period that have 4 
telescopes participating, stable rates (ie no clouds or data acquisition issues) 
and noted as clear by the observers in the shift logs. The data points are 
sub-divided according to the aerosol boundary layer conditions and the
$T_{\mathrm{sky}}$  at zenith for that run, the correlation between warm 
sky temperature, aerosol presence and lowered telescope trigger rate is clearly 
apparent.

\section{Discussion \& Conclusions}
The atmospheric clarity conditions according to lidar and infra-red radiometer
measurements have been presented here. The presence of aerosols in the
atmosphere show up clearly in the lidar returns and also as a clear increase in
$T_\mathrm{sky}$. The data selected here come from a single two 
week period to ensure no seasonal temperature effects can bias the dataset. The
$T_\mathrm{sky}$ will still change somewhat due to the day-to-day ambient 
temperature variation, however this would be expected to produce no more than a 
$\sim20$\% difference -- not the observed $\sim$200\%.
During the most severe periods of aerosol contamination the boundary layer can
be seen to extend to relatively high altitudes. As the production height for 
air shower photons is above these aerosol layers they should act like filters 
only, but since the light of muon ring images (commonly used to determine the
atmosphere's contribution to the systematic uncertainty) develop within these 
layers they will have a distinctly different and more complicated response to 
different boundary layer altitudes. This will be something to examine in future 
work.

In summary, the lidar is extremely useful in determining the presence of aerosol
layers and measuring the transmission profiles, but has limited resolution at 
altitudes $\leq$1\,km and limitations as to when it can be operated; the 
infra-red radiometer is sensitive to the presence or absence of aerosols, 
operates all of the time and will be most sensitive to low altitude aerosols. 
Together they have the potential to quantify atmospheric opacity entirely 
independently of the telescope systematics.

%%%%%%%%%%%%%%%%%%%%%%%%%%%%%%%%%%%%%%%%%%%%%%%%
%% BACKMATTER
%%%%%%%%%%%%%%%%%%%%%%%%%%%%%%%%%%%%%%%%%%%%%%%%

%\begin{theacknowledgments}
%We appreciate the excellent work of the technical support staff in Durham,
%Montpellier and in Namibia in the construction and operation of the equipment.
%\end{theacknowledgments}



\bibliographystyle{aipproc}   % if natbib is available

%\bibliography{sample}

%  \typeout{******************************************}
%  \typeout{** Please run "bibtex \jobname" to optain}
%  \typeout{** the bibliography and then re-run LaTeX}
%  \typeout{** twice to fix the references!}
%  \typeout{******************************************}

%%%%%%%%%%%%%%%%%%%%%%%%%%%%%%%%%%%%%%%%%%%
%% The following lines show an example how to produce a bibliography
%% without the help of the BibTeX program. This could be used instead
%% of the above.
%%%%%%%%%%%%%%%%%%%%%%%%%%%%%%%%%%%%%%%%%%%

\end{document}